\documentclass[aps, prd, reprint, a4paper, nofootinbib, floatfix]{revtex4-2}
\pdfoutput=1
\usepackage[utf8]{inputenc}
\usepackage[english]{babel}
\usepackage[T1]{fontenc}
\usepackage{amsmath}
\usepackage{amsfonts}
\usepackage{amssymb}
\usepackage{amsthm}
\usepackage{makeidx}
\usepackage{graphicx}
\usepackage{color}
\usepackage{xspace}
\usepackage{hyperref}
\hypersetup{citecolor=mpGreen}
\hypersetup{colorlinks=true}
\hypersetup{linkcolor=mpBlue}
\hypersetup{urlcolor=mpBlue}
\definecolor{mpBlue}{RGB}{21, 101, 192}
\definecolor{mpGreen}{RGB}{46, 125, 50}
\DeclareMathOperator{\CC}{\mathbb{C}}
\DeclareMathOperator{\e}{e}
\DeclareMathOperator{\linspan}{span}
\DeclareMathOperator{\Ha}{\mathcal{H}}
\DeclareMathOperator{\Ka}{\mathcal{K}}
\DeclareMathOperator{\bound}{\mathcal{B}}
\DeclareMathOperator{\dens}{\mathcal{D}}
\usepackage{dsfont}
\DeclareMathOperator{\I}{\mathds{1}}
\DeclareMathOperator{\Tr}{Tr}
\DeclareMathOperator{\id}{id}
\DeclareMathOperator{\swap}{SWAP}
\newcommand*{\gme}{\mathrm{GME}}
\newcommand*{\abs}[1]{\left\lvert #1 \right\rvert}
\makeatletter
\newcommand*{\bra}[1]{\langle #1 \@ifnextchar\ket{}{|}}
\newcommand*{\ket}[1]{| #1 \rangle \@ifnextchar\bra{\!}{}}
\makeatother

\newcommand*{\ketbra}[1]{| #1 \rangle \! \langle #1 |}
\newcommand*{\unit}[1]{\,\mathrm{#1}}
\newtheorem{theorem}{Theorem}
\newtheorem{assumption}{Assumption}
\renewcommand{\paragraph}[1]{\addcontentsline{toc}{section}{#1}\emph{#1.}---}

\begin{document}
\title{Existing experiments suffice to indirectly verify the quantum essence of gravity}

\author{Martin Pl\'{a}vala}
\affiliation{Institut für Theoretische Physik, Leibniz Universität Hannover, 30167 Hannover, Germany}

\begin{abstract}
The gravity-mediated entanglement experiments employ concepts from quantum information to argue that if entanglement due to gravitational interaction is observed, then gravity cannot be described by a classical system. However, the proposed experiments remain beyond out current technological capability, with optimistic projections placing the experiment outside of short-term future. Here we argue that current matter-wave interferometers are sufficient to indirectly prove that gravitational interaction creates entanglement between two systems. Specifically, we prove that if we experimentally verify the Schr\"{o}dinger equation for a single delocalized system interacting gravitationally with an external mass, then, under one of two reasonable assumptions, the time evolution of two delocalized systems will lead to gravity-mediated entanglement.
\end{abstract}

\maketitle

\paragraph{Introduction}%
Quantum theory has a long track record of experimentally confirmed predictions, and it is extremely successful theory of electromagnetic, weak, and strong interactions. Yet a unified theory that would include gravity seems to be outside of our reach. Several theories of quantum gravity were proposed \cite{kiefer2012quantum}, and various experiments were carried out \cite{colella1975observation,asenbaum2017phase,overstreet2022observation,panda2024measuring}, but we still lack widely accepted and experimentally tested quantum theory of gravity.

Quantum information was developed to investigate the information-processing capabilities of quantum systems. Quantum information research had an unintended but welcome consequence: it significantly deepened our understanding of quantum theory itself, for example by developing the theory of quantum entanglement \cite{werner1989quantum}. Entanglement is a type of quantum correlation between two systems that is a resource for quantum key distribution \cite{ekert1991quantum}, but also for computational speed-up of quantum computers \cite{jozsa1997entanglement}. A well-known result is that entanglement cannot be created by local operations and classical communication (LOCC) but only via quantum interaction between the systems.

One of the first applications of quantum information to quantum gravity are the recently-proposed experiments to observe gravity-mediated entanglement (GME) \cite{bose2017spin,marletto2017gravitationally}, i.e., to observe the entanglement between two quantum systems created due to their gravitational interaction. These experiments aim to settle the question whether gravity needs to be quantized \cite{tilloy2019does,oppenheim2023time}. While observing GME would rule out some alternative theories of gravity \cite{oppenheim2023postquantum,ludescher2025gravity}, the exact implications of the observation are still unclear, their conclusions were heavily scrutinized \cite{hall2018two,anastopoulos2021gravitational,fragkos2022inference,doner2022gravitational, marshman2020locality,christodoulou2022gravity,danielson2022gravitationally,bose2022mechanism, christodoulou2023locally,marchese2025newton,trillo2025diosi,tibau2025bose}, and recently intensively discussed \cite{aziz2025classical, marletto2025classical1, lin2025newtonian, diosi2025no, di2025simple, marletto2025classical2, schneider2025demonstration,sienicki2025comment}.

But the main problem is that even optimistic estimates place the experiment outside of short-term future \cite{aspelmeyer2022zeh}. The GME experiment would require us to prepare two quantum systems delocalized in space, that is, both systems need to be in superposition of two spatialy distinct states. A considerably simpler experiments involve only a single quantum system in spatial superposition interacting with other localized systems. Quantum systems in spatial superpositions can be prepared using matter-wave interferometers, current experimental setups are already capable of delocalizing quantum system over half meter in distance \cite{kovachy2015quantum} or to keep the systems in quantum state for over one minute \cite{panda2024coherence}. Such setup was recently used to observe the gravitational Aharonov-Bohm effect \cite{overstreet2022observation}, and it was recently proposed to verify the Schr\"{o}dinger equation for a single delocalized particle in a quantum regime beyond the classical limit with such setup \cite{plavala2025probing}.

In this paper we will use arguments employing the machinery of quantum information to prove that if we experimentally verify the Schr\"{o}dinger equation for a single delocalized system according to the recent proposal \cite{plavala2025probing}, then, under one of two reasonable assumptions, entanglement must be created when two delocalized systems interact via gravity. Thus existence of GME could be indirectly confirmed by today's matter-wave interferometers.

\begin{figure}
\includegraphics[width=\linewidth]{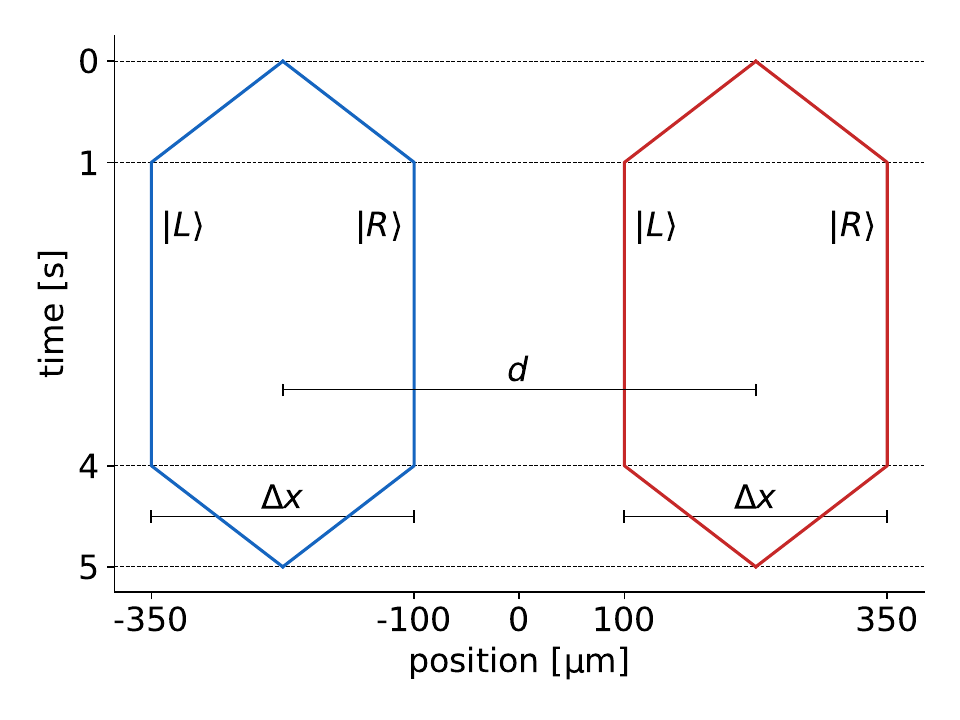}
\caption{The spacetime diagram depicting two mass interferometers (blue and red) separated by a distance $d = \unit{450 \mu{}m}$, the values are based on \cite{bose2017spin}. From time $t=0$ to time $t=1 \unit{s}$ magnetic field or laser pulses are used to coherently split the momentum of both particles, thus yielding two arms, $|L\rangle$ and $|R\rangle$, of the respective interferometers separated by the distance $\Delta x = \unit{250 \mu{}m}$. Then the arms are kept at constant separations and their gravitational interaction will (according to the Schr\"{o}dinger equation) cause relative phaseshifts between the four combinations of the arms, entangling the two systems. The experiment concludes by applying the interferometric sequence in reverse, recombining the arms, and enabling to observe the gravity-mediated entanglement (GME). \label{fig:two}}
\end{figure}

\begin{figure}
\includegraphics[width=\linewidth]{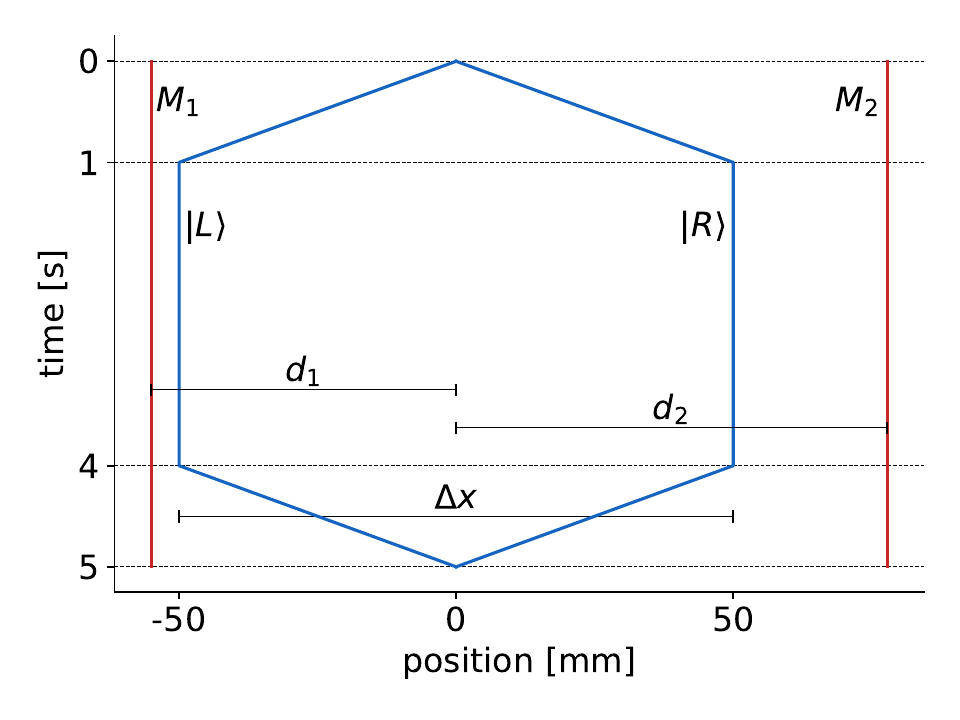}
\caption{The spacetime diagram depicting a mass interferometer (blue) and macroscopic test masses $M_1$ and $M_2 = 2 M_1$ (red) at distances $d_1 = 55 \unit{mm}$ and $d_2 \approx \unit{78 \unit{mm}}$ from the center of mass of the interferometer, the values are based on \cite{plavala2025probing}. From time $t = 0$ to $t = 1 \unit{s}$ magnetic field or laser pulses are used to coherently split the momentum of the particle, thus yielding two arms, $|L\rangle$ and $|R\rangle$, separated by $\Delta x = 10 \unit{cm}$. Then the arms are kept at constant separation. The external masses are positioned so that the phase shift due to the classical limit vanishes and thus any effects will be only due to quantum effects of gravity. The experiment concludes by applying the interferometric sequence in reverse and measuring the quantum effects of gravity. \label{fig:one}}
\end{figure}

\paragraph{Unknown time evolution}%
Since quantum gravity is an unknown theory, we should use a general framework, such as the general probabilistic theories \cite{plavala2022operational,plavala2023generalized,plavala2023general} to model the unknown theory. Using such general framework, it was argued in \cite{marchese2025newton,plavala2025probing} that the only unknown term in the low energy experiments is the time evolution since the initial states and final measurements are always implemented using electromagnetic interactions.

Consider a gravitational interaction of two massive particles and let $\Phi_t$ denote the unknown time evolution, then $\Phi_t: \dens(\Ha \otimes \Ha) \to \dens(\Ha \otimes \Ha)$ is a linear map that maps the density matrix of the initial state to a density matrix describing the final state, here we use $\dens(\Ha)$ to denote a set of density matrices on a Hilbert space $\Ha$. We assume that the time evolution is linear, an examples of nonlinear time evolution would be due to the Schr\"{o}dinger-Newton equation \cite{ruffini1969systems}: such nonlinearities may arise due to effects of general relativity \cite{amelino2001status}, and they can be ruled out using contextuality \cite{delgado2025ruling}.

Even if we consider the time evolution $\Phi_t$ implementing the mutual gravitational interaction of two systems to be unknown, it must be physically well defined. It cannot predict the statistics of any experiment to be described by negative probabilities since these are meaningless. That is, $\Phi_t$ must be a completely positive quantum channel, or at least positive quantum channel. $\Phi_t$ is a positive quantum channel if for arbitrary initial state of the two systems $\rho \in \dens(\Ha_2 \otimes \Ha_2)$ we have that the final state of the system after the experiment, denoted $\Phi_t(\rho)$, is a well-defined quantum state, $\Phi_t(\rho) \in \dens(\Ha_2 \otimes \Ha_2)$. A stronger yet more common requirement is that $\Phi_t$ is completely positive, that is, even if the initial state would be entangled to arbitrary other system, then the final state of the experiment is a well-defined quantum state.
It is debatable whether we should consider the unknown time evolution $\Phi_t$ only positive or completely positive, since complete positivity essentially assumes that we can make a part of large system evolve in time while another part of it remains frozen in time.

\paragraph{Proposed experiments}%
The proposed experiments are based on matter-wave interferometers, where massive quantum system is in the superposition of following two possible paths. Thus each quantum system in question is effectively described by a qubit Hilbert space $\Ha_2$ spanned by two orthogonal vectors $\ket{L}$ and $\ket{R}$, $\bra{L}\ket{R} = 0$, corresponding to the left and right paths respectively. In the GME experiments we initialize both systems in equal superposition of both paths, $\ket{+} = (\ket{L} + \ket{R}) / \sqrt{2}$, and we aim to determine whether the final state $\Phi_t(\ketbra{+} \otimes \ketbra{+})$ is entangled or not, see Fig.~\ref{fig:two}. According to the Schr\"{o}dinger equation, the gravitational interaction is described by the Hamiltonian
\begin{equation} \label{eq:timeEvo-Hamiltonian}
\hat{H} = - \frac{G m_1 m_2}{\abs{\hat{x} - \hat{y}}}
\end{equation}
where $m_1, m_2$ and  $\hat{x}, \hat{y}$ denote the masses and position operators of the first and second system respectively. Such interaction does entangle the two systems in question. Thus any possibility of not observing the GME implies that the Schr\"{o}dinger equation has to be modified.

An experiment to verify the Schr\"{o}dinger equation for single delocalized system was recently proposed in \cite{plavala2025probing}: while verifying the Schr\"{o}dinger equation is the expected outcome consistent with GME existing, finding deviations from the Schr\"{o}dinger equation would not immediately imply that GME does not exist. Key aspects of this experiment is that it operates in a quantum regime beyond the classical limit, unlike previous experiments \cite{colella1975observation}, and that it is feasible with today's matter-wave interferometers. The experiment includes a delocalized system and one or more external masses that source the gravitational field, see Fig.~\ref{fig:one}. The external masses present in the experiment have a well defined position and thus can be described by the (formal) eigenstates of the position operator. Hence we can treat this experiments in the same framework as the GME experiment.

Consider for simplicity just one external mass. The delocalized system is still described by the qubit Hilbert space $\Ha_2$, and we assign a quantum state $\ketbra{x}$ localized at position $x$ to the localized system. While we cannot predict the outcome of the proposed experiments with certainty, the Schr\"{o}dinger equation is the most natural candidate and thus we should expect that the experiments will conclude that for arbitrary $\ket{\psi} \in \Ha_2$ we must have
\begin{equation}
\Phi_t(\ketbra{\psi} \otimes \ketbra{x}) = \e^{- \frac{it\hat{H}}{\hbar}} \ketbra{\psi} \otimes \ketbra{x} \e^{\frac{it\hat{H}}{\hbar}}, \label{eq:timeEvo-Phit-psiXX}
\end{equation}
where, according to the Schr\"{o}dinger equation, the interaction Hamiltonian is yet again given by \eqref{eq:timeEvo-Hamiltonian}. Note that even if \eqref{eq:timeEvo-Phit-psiXX} is experimentally confirmed, this does not immediately imply that $\Phi_t(\ketbra{\varphi}) = \e^{- \frac{it\hat{H}}{\hbar}} \ketbra{\varphi} \e^{\frac{it\hat{H}}{\hbar}}$ for arbitrary $\ket{\varphi} \in \Ha_2 \otimes \Ha_2$, since the second systems in \eqref{eq:timeEvo-Phit-psiXX} is restricted to the (formal) eigenbasis of position. This is a key difference between the verification and GME experiments.

Are the outcomes of these two experiment depicted in Fig.~\ref{fig:two} and \ref{fig:one} independent? As already pointed out, Schr\"{o}dinger equation does predict GME and so not observing GME does imply a necessary modification of the Schr\"{o}dinger equation. But do we learn anything about the existence of GME from verifying the Schr\"{o}dinger equation? These experiments do operate in different parameter regimes and so it is not straightforward to draw conclusions.

\paragraph{Assumptions}
We will outline two assumptions that will be used to draw conclusions about the existence of GME from the outcome of the verification experiments depicted in Fig.~\ref{fig:one}. Only one of the two assumptions is necessary to indirectly prove the existence of GME. Thus any theory of gravity that does not predict GME and is consistent with \eqref{eq:timeEvo-Phit-psiXX} has to violate both assumptions.

\begin{assumption} \label{ass:delocalized}
A massive body with weight of $M \approx 20 \unit{g}$ can be delocalized over arbitrary small distance $\Delta y$ with negligible decoherence and the time evolution is symmetric, meaning that
\begin{equation} \label{eq:assumption1-symmetry}
\swap \circ \; \Phi_t \circ \swap = \Phi_t,
\end{equation}
where $\swap$ denotes the superoperator that exchanges systems, $\swap(\hat{X} \otimes \hat{Y}) = \hat{Y} \otimes \hat{X}$ for arbitrary operators $\hat{X}, \hat{Y}$, and $\circ$ denotes the composition of maps.
\end{assumption}
The requirement that the time evolution is symmetric is natural: it is reminiscent of the principle of action and reaction, as it states that exchanging the particles before or after an interaction leads to the same result. The assumption that a body with weight of $M \approx 20 \unit{g}$ can be delocalized could be in principle tested. While spatial superpositions of large molecules were observed \cite{fein2019quantum}, it is unlikely that Assumption~\ref{ass:delocalized} would be experimentally testable soon. We will provide a qualitative and analytic proof that Assumption~\ref{ass:delocalized} and a verification of the Schr\"{o}dinger equation for single delocalized system implies existence of GME.

\begin{assumption} \label{ass:massProduct}
The interaction as described by the unknown linear map $\Phi_t$ depends only on the product of the masses $m_1 m_2$ of the two massive particles. Or, more generally, for $\lambda > 0$ we have that, within reasonable mass range, the interaction between particles of mass $\lambda m_1$ and $m_2$ is the same as between particles of mass $m_1$ and $\lambda m_2$.
\end{assumption}
This assumption is related to the equivalence principle and the universality of free fall, which were recently experimentally tested \cite{schlippert2014quantum,hartwig2015testing,rosi2017quantum,bhadra2024test}. Also experiments testing gravitational field of smaller and smaller objects were carried out and no discrepancy was observed so far \cite{tan2020improvement,westphal2021measurement,fuchs2024measuring,panda2024measuring}. We will provide a quantitative and numerical proof that Assumption~\ref{ass:massProduct} and a verification of the Schr\"{o}dinger equation for single delocalized system implies existence of GME.



\paragraph{Assumption~\ref{ass:delocalized} and a qualitative proof of GME}%
Here we prove that if \eqref{eq:timeEvo-Phit-psiXX} holds exactly, if the time evolution $\Phi_t$ is completely positive, and if Assumption~\ref{ass:delocalized} holds, then GME exists. This proof is only qualitative, since we need to assume that \eqref{eq:timeEvo-Phit-psiXX} holds exactly, or that any decoherence caused by gravity is so small that it can be ignored. This is, of course, somewhat unrealistic, since any experiment verifying the Schr\"{o}dinger equation will at best give an upper bound on the decoherence caused by gravity. We show later that if we replace Assumption~\ref{ass:delocalized} by Assumption~\ref{ass:massProduct}, then we can obtain quantitative proof taking into account only the upper bound coming from the experiment.

\begin{theorem}
Assume that \eqref{eq:timeEvo-Phit-psiXX} holds exactly for a single delocalized microscopic particle (such as Caesium or Rubidium atom) in the gravitational field of an external mass in the mass range $M \approx 20 \unit{g}$, assume that the time evolution $\Phi_t$ is completely positive, and that Assumption~\ref{ass:delocalized} holds. Then gravity-mediated entanglement (GME) exists.
\end{theorem}
\begin{proof}
In the proposed experiment to verify the Schr\"{o}dinger equation, the localized mass generating the gravitational field $M = 20 \unit{g}$. According to Assumption~\ref{ass:delocalized} such mass can be delocalized and thus it can be put in arbitrary superposition of two spatially-localized states $\ket{L}$ and $\ket{R}$ that are separated by $\Delta y$. Let $\ket{\psi} \in \Ha_2$ be an arbitrary state of the matter-wave interferometer and let $y \in \{L, R\}$ be such that $\ket{y}$ is a localized state of the other atom interferometer. Then, using the symmetry of the time evolution coming from Assumption~\ref{ass:delocalized}, we get $\Phi_t(\ketbra{y} \otimes \ketbra{\psi}) = \swap(\Phi_t(\ketbra{\psi} \otimes \ketbra{y}))$. Using \eqref{eq:timeEvo-Phit-psiXX}, we get
\begin{equation}
\begin{split}
\Phi_t(\ketbra{y} \otimes \ketbra{\psi}) &= \swap(\e^{- \frac{it\hat{H}}{\hbar}} \ketbra{\psi} \otimes \ketbra{y} \e^{- \frac{it\hat{H}}{\hbar}}) \\
&= \e^{- \frac{it\hat{H}}{\hbar}} \ketbra{y} \otimes \ketbra{\psi} \e^{\frac{it\hat{H}}{\hbar}}, \label{eq:assumption1-Phit-YYpsi}
\end{split}
\end{equation}
where we used the symmetry of the interaction Hamiltonian \eqref{eq:timeEvo-Hamiltonian} in the last step. We thus get that the Schr\"{o}dinger equation holds if just one of the particles is delocalized, no matter which one.

The rest of the proof is based on expressing the unknown time evolution $\Phi_t$ via its Choi matrix $J(\Phi_t)$ \cite{jamiolkowski1972linear,choi1975completely}. We show that $J(\Phi_t)$ is uniquely determined by \eqref{eq:timeEvo-Phit-psiXX} and \eqref{eq:assumption1-Phit-YYpsi}, and complete positivity of $\Phi_t$. This implies that the time evolution is given by the Schr\"{o}dinger equation even for two delocalized systems, which is known to give rise to GME. We relegate this part of the proof to Appendix~\ref{appendix:analytic}.
\end{proof}

\paragraph{Assumption~\ref{ass:massProduct} and a quantitative proof of GME}%
Here we present a different proof that GME exists: given that Assumption~\ref{ass:massProduct} holds, we will be able to numerically investigate an equivalent version of the original GME experiment and show that it will result in an entangled state, and thus that GME exist. Thanks to the use of numerical methods, we will not have to require that \eqref{eq:timeEvo-Phit-psiXX} holds exactly, but it will be sufficient to assume that the decoherence due to gravity is upper bounded within the precision available in upcoming experiments. The form of the upper bound will be very general: we will only put constraint on the effective decay of the off-diagonal elements of the density matrix. This means that, unlike previous research in this direction \cite{kryhin2025distinguishable,trillo2025diosi}, we do not have to consider a specific alternative theory of gravity, but we will formulate our result directly in terms of experimentally-observable quantities.

To model the decoherence, instead of \eqref{eq:timeEvo-Phit-psiXX}, we will assume that the off-diagonal elements of the density matrix are suppressed by gravity. That is, we will assume that $\Phi_t(\ket{L}\bra{R} \otimes \ketbra{x}) = \lambda_t \e^{- \frac{it\hat{H}}{\hbar}} \ket{L}\bra{R} \otimes \ketbra{x} \e^{\frac{it\hat{H}}{\hbar}}$ and analogically for the hermitian conjugate. $\lambda_t$ quantifies the decoherence caused by gravity, here $\lambda_t = 1$ implies no decoherence, while $\lambda_t = 0$ is complete decoherence where the off-diagonal terms of the density matrix vanish. $\lambda_t$ will be lower bounded by the experiment to verify Schr\"{o}dinger equation: while the experiment cannot certify that $\lambda_t = 1$ as predicted by the Schr\"{o}dinger equation, it can only give us an upped bound on decoherence, which is a number $\lambda_0$ such that $\lambda_t \geq \lambda_0$.\footnote{The measured quantity $\lambda_0$ is an upper bound on the decoherence since it limits the potential decoherence. But we have $\lambda_t \geq \lambda_0$, so $\lambda_0$ is numerically a lower bound on the respective parameter. This stems from $\lambda_t = 1$ corresponding to no decoherence and unitary time evolution.} As we will show, today's matter-wave interferometers are already capable of sufficiently upper bounding the decoherence, that is of measuring $\lambda_0$ with sufficient precision, to numerically certify the existence of GME.

Note that we will still assume $\Phi_t(\ketbra{x} \otimes \ketbra{y}) = \e^{- \frac{it\hat{H}}{\hbar}} \ketbra{x} \otimes \ketbra{y} \e^{\frac{it\hat{H}}{\hbar}} = \ketbra{x} \otimes \ketbra{y}$ since time evolution of localized particles is trivial as we are neglecting the relative movement.
\begin{theorem}
Assume that the time evolution $\Phi_t$ is positive, and that Assumption~\ref{ass:massProduct} holds. Then a sufficiently precise experiment to verify the Schr\"{o}dinger equation can prove that GME exist. The necessary precision is available in today's matter-wave interferometers.
\end{theorem}
\begin{proof}
Due to the numerical nature of the proof, the exact result depends on the specifics of a given experimental setup to verify the Schr\"{o}dinger equation. Here we will analyze one concrete case that demonstrates how to find the lower bound for decoherence sufficient to prove the existence of GME. Analogical calculations are straightforward to carry out for other specific setups.

We will assume that the Schr\"{o}dinger equation was experimentally verified for a single Ceasium atom with mass $m = 2 \cdot 10^{-25} \unit{kg}$ and a Tungsten sphere with mass $M = 10 \unit{kg}$ and radius sligtly less than $5 \unit{cm}$. We will assume that the distance from the surface of the sphere to one of the arms of the matter-wave interferometer is $3 \unit{mm}$, as proposed in \cite{plavala2025probing}. The result of such experiment will be the verification of the Schr\"{o}dinger equation with an upper bound $\lambda_0$ such that $\lambda_t \geq \lambda_0$ for the given time of measurement. Current matter-wave interferometers have phase sensitivity of about $50 \mu\unit{rad}$ in differential measurements, which translates to sensitivity of less than $1.6 \cdot 10^{-4}$ in measuring $\lambda_0$.

We will now use Assumption~\ref{ass:massProduct}. The consequence of this assumption is that the interaction will be the same for two mesoscopic particles with mass $m_\gme$ as long as $m_\gme^2 = mM$; in this case we get approximately $m_\gme = 1.5 \cdot 10^{-12} \unit{kg}$. We also have to take into account the geometry of the experiment: in the verification experiment, the distance between one of the arms of the matter-wave interferometer and center of mass of the localized tungsten sphere was $5.3 \unit{cm}$. Hence we can only assume that gravity was tested with the closes separation of $5.3 \unit{cm}$ and no less even for mesoscopic particles.

It thus follows that, due to Assumption~\ref{ass:massProduct}, we have
\begin{align}
\Phi_t(\ket{L}\bra{R} \otimes \ketbra{x}) &= \mu^{(1)}_{x,t} \e^{- \frac{it\hat{H}}{\hbar}} \ket{L}\bra{R} \otimes \ketbra{x} \e^{\frac{it\hat{H}}{\hbar}}, \label{eq:assumption2-Phit-LRXX} \\
\Phi_t(\ketbra{y} \otimes \ket{L}\bra{R}) &= \mu^{(2)}_{y,t} \e^{- \frac{it\hat{H}}{\hbar}} \ketbra{y} \otimes \ket{L}\bra{R} \e^{\frac{it\hat{H}}{\hbar}}, \label{eq:assumption2-Phit-YYLR}
\end{align}
for $\mu^{(1)}_{x,t} \geq \lambda_0$ and $\mu^{(2)}_{y,t} \geq \lambda_0$, and analogically for the hermitian conjugate. This holds for two mesoscopic particles with mass $m_\gme$ separated by at least $5.3 \unit{cm}$. \eqref{eq:assumption2-Phit-LRXX} is a quantitative version of \eqref{eq:timeEvo-Phit-psiXX}, \eqref{eq:assumption2-Phit-YYLR} follows from \eqref{eq:assumption2-Phit-LRXX} since the particles are identitcal.\footnote{We only need Assumption~\ref{ass:massProduct} to derive \eqref{eq:assumption2-Phit-LRXX} and \eqref{eq:assumption2-Phit-YYLR} for two mesoscopic particles. It thus follows that if, instead of microscopic and macroscopic particle, one could verify the Schr\"{o}dinger equation for two mesoscopic particles, then Assumption~\ref{ass:massProduct} would not be necessary. Such verification experiment with two mesoscopic particles may be still easier than the GME experiment since it requires only one of the particles to be delocalized.}

The question now is: is there a time evolution represented by a linear map $\Phi_t$, such that $\Phi_t$ is positive, \eqref{eq:assumption2-Phit-LRXX} and \eqref{eq:assumption2-Phit-YYLR} hold, and the final state of the GME experiment $\Phi_t(\ketbra{+} \otimes \ketbra{+})$ is separable? We will answer this question using semidefinite programming \cite{boyd2004convex}.

Since our system consists of two qubits, deciding whether the final state of the GME experiment $\Phi_t(\ketbra{+}\otimes\ketbra{+})$ is entangled or not can be done via the positive partial transpose (PPT) criterion \cite{peres1996separability,horodecki2001separability}. Thus, given an experimentally observed value of $\lambda_0$, we want to maximize the lowest eigenvalue of the partial transpose of $\Phi_t(\ketbra{+}\otimes\ketbra{+})$ over all possible positive linear maps $\Phi_t$ satisfying \eqref{eq:assumption2-Phit-LRXX} and \eqref{eq:assumption2-Phit-YYLR}. If the resulting maximum is negative, then $\Phi_t(\ketbra{+}\otimes\ketbra{+})$ is entangled for all possible $\Phi_t$.

\begin{figure}
\includegraphics[width=\linewidth]{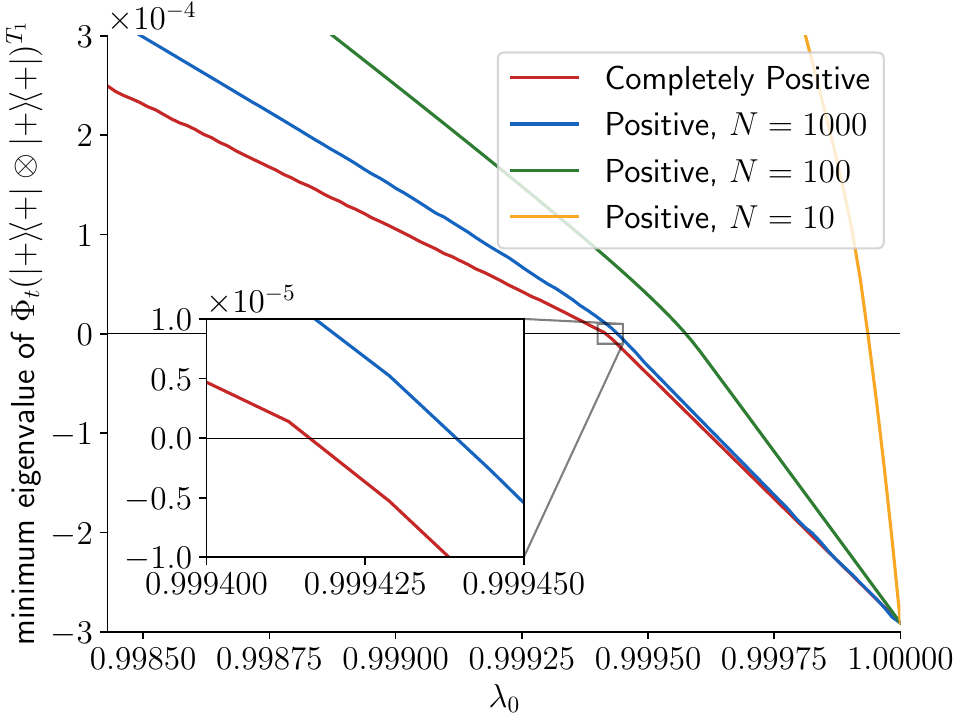}
\caption{The plot of the smallest eigenvalue of the partial transpose of the final state of the GME experiment $\Phi_t(\ketbra{+} \otimes \ketbra{+})$ as a function of the decoherence upper bound $\lambda_0$. If the smallest eigenvalue is negative, then the final state of the GME experiment $\Phi_t(\ketbra{+} \otimes \ketbra{+})$ is entangled for all possible time evolutions $\Phi_t$ corresponding to the given value of $\lambda_0$. The red line is for completely positive time evolution $\Phi_t$, the blue, green, and yellow lines are outer approximations of positive time evolution $\Phi_t$ by enforcing positivity on $1000$, $100$, and $10$ randomly generated pure states. \label{fig:assumption2-numerics}}
\end{figure}

This can be readily implemented by once again representing $\Phi_t$ by its Choi matrix, the only problem is the requirement that $\Phi_t$ is positive as there is no good method of implementing such constraint in semidefinite programming. We relax this condition and we only require that for a set of $N$ randomly selected initial pure states $\{ \rho_i \}_{i=1}^N$, the final states $\Phi_t(\rho_i)$ are valid physical states. While this is strictly weaker condition than positivity, we found that for $N = 1000$ we get a numerical certificate that the unknown positive time evolution $\Phi_t$ will create GME. The results of the numerical calculations are plotted in Fig.~\ref{fig:assumption2-numerics}, where it is shown that for both positive and completely positive time evolution it is sufficient to certify $\lambda_0 \approx 0.9995$, which is within the sensitivity of the current matter-wave interferometers. The detailed description of the numeric methods is provided in Appendix~\ref{appendix:numeric}. The implementation of the semidefinite program using the Python package PICOS \cite{PICOS} is publicly available \cite{MP_github}.
\end{proof}

\paragraph{Conclusions}%
We have, given the clearly outlined assumptions, proven that current matter-wave interferometers are sufficient for indirectly proving the existence of GME. The key conclusions we arrive to are that gravity is likely quantum, that we already have first indications that this is the case \cite{overstreet2022observation}, and that we will have experimental proof considerably sooner than previously expected. But, more importantly, the main bottleneck in conclusively determining whether gravity is quantum is on the side of theory rather than experiment: while the upcoming experiments will indirectly prove the existence of GME, we are still lacking consensus on what are the exact implications of GME for gravity.

The most immediate follow-up work is to figure out the exact conclusions of the existence of GME. This will require us to reconcile relativistic locality as used in quantum field theory with subsystem locality as used in quantum information \cite{di2023relativistic} and it is likely that we will require generalized framework \cite{plavala2022operational,plavala2023generalized} for quantum field theories in order to draw conclusive verdict about the implications of GME. It is also an open question whether one can relax our assumptions: for example in the axiomatization program for quantum theory, the initial axioms \cite{hardy2001quantum} were later simplified and replaced by more natural assumptions \cite{chiribella2011informational}. An additional open question worth mentioning is pertaining to the positive time evolution: it would be useful to clarify whether we need to consider positive time evolution, or whether one can present sufficient arguments that the time evolution can be considered completely positive in the non-relativistic regime. We leave these questions open for future work.

\paragraph{Acknowledgments}
I am thankful to Ren\'{e} Schwonnek, Leevi Lepp\"{a}j\"{a}rvi, Guillermo Alejandro P\'{e}rez Lobato, Michael Werner, Peter Asenbaum, and Alexandre Gauguet for discussions.
This work was supported by the Nieders\"{a}chsisches Ministerium f\"{u}r Wissenschaft und Kultur.

\bibliography{citations}

\onecolumngrid
\appendix

\section{Analytic proof that Schr\"{o}dinger equation for single delocalized system and complete positivity of time evolution implies gravity-mediated entanglement} \label{appendix:analytic}
Let $\Ha$ be a finite-dimensional complex Hilbert space, then we will use $\bound(\Ha)$ to denote the operators on $\Ha$. An operator $A \in \bound(\Ha)$ is positive semidefinite, denoted $A \geq 0$, if $A$ is hermitian and all of its eigenvalues are nonnegative. $\dens(\Ha)$ will denote the set of density matrices on $\Ha$, so $\rho \in \dens(\Ha)$ if and only if $\rho \geq 0$ and $\Tr(\rho) = 1$.

The system in question is represented as two qubits, where each qubit corresponds to a single quantum system being in the left path or right path. Thus the underlying Hilbert space is of the form $\Ha_2 \otimes \Ha_2$, where $\Ha_2 = \linspan(\ket{L}, \ket{R})$, where $\ket{L}$ and $\ket{R}$ are the vectors representing the system being localized in the left or right path, $\bra{L}\ket{R} = 0$. We will also use $\hat{x}$ and $\hat{y}$ to denote the position operators of the first and second system respectively, $x_L, x_R$ will denote the position of the left and right arms of the first interferometer and $y_L, y_R$ will denote the position of the left and right arms of the second interferometer, respectively, that is
\begin{align}
&\hat{x} \ket{L_1} = x_L \ket{L_1},
&&\hat{y} \ket{L_2} = y_L \ket{L_2}, \\
&\hat{x} \ket{R_1} = x_R \ket{R_1},
&&\hat{y} \ket{R_2} = y_R \ket{R_2},
\end{align}
where we have explicitly distinguished the Hilbert space of the first particle $\linspan(\ket{L_1}, \ket{R_1})$ and of the second particle $\linspan(\ket{L_2}, \ket{R_2})$. In the following this distinction will be clear from the position of the vector in the tensor product.

In general we will treat the time evolution as an unknown linear map $\Phi_t$ that takes a density matrix of the initial state and outputs the density matrix of the final state at time $t$. This will enable us to use the machinery of quantum information, and is also necessary as we want to investigate the case of general time evolution that can cause decoherence. Thus in general the time evolution $\Phi_t$ maps density matrices to density matrices,
\begin{equation}
\Phi_t: \dens(\Ha_2 \otimes \Ha_2) \to \dens(\Ha_2 \otimes \Ha_2).
\end{equation}
Moreover we require $\Phi_t$ to be trace preserving, which follows from the requirement that probabilities for arbitrary measurement sum up to one. Mathematically, $\Phi_t$ is trace preserving if and only if
\begin{equation}
\Tr(\Phi_t(A)) = \Tr(A)
\end{equation}
for any $A \in \bound(\Ha_2 \otimes \Ha_2)$. In this section we will additionally require that $\Phi_t$ is completely positive, meaning that the final state must be a valid density matrix, even if the system would be entangled with arbitrary other system. Mathematically, complete positivity is expressed as follows: for arbitrary Hilbert space $\Ka$ and for arbitrary $\rho \in \dens((\Ha_2 \otimes \Ha_2) \otimes \Ka)$ we must have that
\begin{equation}
(\Phi_t \otimes \id)(\rho) \geq 0,
\end{equation}
that is, that $(\Phi_t \otimes \id)(\rho)$ is positive semidefinite and hence a valid density matrix.

Let
\begin{equation}
\hat{H} = - \frac{G m_1 m_2}{\abs{\hat{x} - \hat{y}}}
\end{equation}
be the Hamiltonian of the two quantum systems interacting via gravity. We are neglecting the kinetic part of the Hamiltonian, i.e., the contribution from the kinetic energy of the systems, since the displacement due to mutual gravitational interaction is negligible and can be either taking into account by using the classical paths of the systems \cite{overstreet2021physically} or counteracted in lattice interferometers \cite{panda2024coherence}.

We argued in the main text that, as a result of the verification experiment and the respective assumption, we get that for arbitrary $\ket{\psi} \in \Ha_2$ it holds that
\begin{align}
\Phi_t(\ketbra{\psi} \otimes \ketbra{y}) &= \e^{- \frac{it\hat{H}}{\hbar}} \ketbra{\psi} \otimes \ketbra{y} \e^{\frac{it\hat{H}}{\hbar}}, \label{eq:analytic-Phit-psiYY} \\
\Phi_t(\ketbra{x} \otimes \ketbra{\psi}) &= \e^{- \frac{it\hat{H}}{\hbar}} \ketbra{x} \otimes \ketbra{\psi} \e^{\frac{it\hat{H}}{\hbar}}, \label{eq:analytic-Phit-XXpsi}
\end{align}
where $x,y \in \{L, R\}$ index the two arms of the interferometers. The Hamiltonian is diagonal in the which path basis, we have
\begin{equation}
\hat{H} = - G m_1 m_2 \left( \frac{\ketbra{LL}}{\abs{x_L - y_L}} + \frac{\ketbra{LR}}{\abs{x_L - y_R}} + \frac{\ketbra{RL}}{\abs{x_R - y_L}} + \frac{\ketbra{RR}}{\abs{x_R - y_R}} \right).
\end{equation}
where $\ketbra{xy} = \ketbra{x} \otimes \ketbra{y}$; we will switch between the two notations as appropriate. This immediately yields that we have
\begin{equation}
\e^{- \frac{it\hat{H}}{\hbar}} = \e^{i \varphi_{LL}} \ketbra{LL} + \e^{i \varphi_{LR}} \ketbra{LR} + \e^{i \varphi_{RL}} \ketbra{RL} + \e^{i \varphi_{RR}} \ketbra{RR}
\end{equation}
where $\varphi_{LL}, \varphi_{LR}, \varphi_{RL}, \varphi_{RR}$ are the corresponding phases, e.g., $\varphi_{LL} = \frac{i G m^2 t}{\hbar \abs{x_L - y_L}}$. We get
\begin{align}
\Phi_t(\ketbra{LL}) &= \e^{- \frac{it\hat{H}}{\hbar}} \ketbra{LL} \e^{\frac{it\hat{H}}{\hbar}} = \ketbra{LL} \label{eq:analytic-Phit-LL} \\
\Phi_t(\ketbra{LR}) &= \e^{- \frac{it\hat{H}}{\hbar}} \ketbra{LR} \e^{\frac{it\hat{H}}{\hbar}} = \ketbra{LR} \label{eq:analytic-Phit-LR} \\
\Phi_t(\ketbra{RL}) &= \e^{- \frac{it\hat{H}}{\hbar}} \ketbra{RL} \e^{\frac{it\hat{H}}{\hbar}} = \ketbra{RL} \label{eq:analytic-Phit-RL} \\
\Phi_t(\ketbra{RR}) &= \e^{- \frac{it\hat{H}}{\hbar}} \ketbra{RR} \e^{\frac{it\hat{H}}{\hbar}} = \ketbra{RR} \label{eq:analytic-Phit-RR}
\end{align}
since the phases cancel each other.

We will use \eqref{eq:analytic-Phit-psiYY} and \eqref{eq:analytic-Phit-XXpsi} to express $\Phi_t(\ket{L}\bra{R} \otimes \ketbra{L})$ using the unitary $\e^{- \frac{it\hat{H}}{\hbar}}$. To do so we will need to express $\ket{L}\bra{R}$ using terms of the form $\ketbra{\psi}$, which is done with the help of the eigenbases of the Pauli $X$ and $Y$ matrices. Let
\begin{align}
\ket{+} &= \frac{1}{\sqrt{2}} (\ket{L} + \ket{R}), \\
\ket{-} &= \frac{1}{\sqrt{2}} (\ket{L} - \ket{R}), \\
\ket{\uparrow} &= \frac{1}{\sqrt{2}} (\ket{L} + i\ket{R}), \\
\ket{\downarrow} &= \frac{1}{\sqrt{2}} (\ket{L} - i\ket{R}),
\end{align}
then we have
\begin{equation} \label{eq:analytic-LRdecomposition}
\ket{L}\bra{R} = \frac{1}{2} ( \ketbra{+} - \ketbra{-} + i (\ketbra{\uparrow} - \ketbra{\downarrow}) ).
\end{equation}
It thus follows from the linearity of $\Phi_t$ that
\begin{equation}
\begin{split}
\Phi_t(\ket{L}\bra{R} \otimes \ketbra{L}) &= \frac{1}{2} ( \Phi_t(\ketbra{+}\otimes \ketbra{L}) - \Phi_t(\ketbra{-}\otimes \ketbra{L}) + i \Phi_t(\ketbra{\uparrow}\otimes \ketbra{L}) - i \Phi_t(\ketbra{\downarrow}\otimes \ketbra{L}) ) \\
&= \frac{1}{2} ( \e^{- \frac{it\hat{H}}{\hbar}} \ketbra{+}\otimes \ketbra{L} \e^{\frac{it\hat{H}}{\hbar}} - \e^{- \frac{it\hat{H}}{\hbar}} \ketbra{-}\otimes \ketbra{L} \e^{\frac{it\hat{H}}{\hbar}} \\
&+ i \e^{- \frac{it\hat{H}}{\hbar}} \ketbra{\uparrow}\otimes \ketbra{L} \e^{\frac{it\hat{H}}{\hbar}} - i \e^{- \frac{it\hat{H}}{\hbar}} \ketbra{\downarrow}\otimes \ketbra{L} \e^{\frac{it\hat{H}}{\hbar}} ) \\
&= \e^{- \frac{it\hat{H}}{\hbar}} \ket{L}\bra{R} \otimes \ketbra{L} \e^{\frac{it\hat{H}}{\hbar}}
\end{split}
\end{equation}
Where we used \eqref{eq:analytic-LRdecomposition} twice and \eqref{eq:analytic-Phit-psiYY}. Repeating the same calculations for all the other off-diagonal elements we get
\begin{align}
\Phi_t(\ket{L}\bra{R} \otimes \ketbra{L}) &= \e^{- \frac{it\hat{H}}{\hbar}} \ket{L}\bra{R} \otimes \ketbra{L} \e^{\frac{it\hat{H}}{\hbar}}, \label{eq:analytic-Phit-LRLL} \\
\Phi_t(\ket{R}\bra{L} \otimes \ketbra{L}) &= \e^{- \frac{it\hat{H}}{\hbar}} \ket{R}\bra{L} \otimes \ketbra{L} \e^{\frac{it\hat{H}}{\hbar}}, \\
\Phi_t(\ket{L}\bra{R} \otimes \ketbra{R}) &= \e^{- \frac{it\hat{H}}{\hbar}} \ket{L}\bra{R} \otimes \ketbra{R} \e^{\frac{it\hat{H}}{\hbar}}, \\
\Phi_t(\ket{R}\bra{L} \otimes \ketbra{R}) &= \e^{- \frac{it\hat{H}}{\hbar}} \ket{R}\bra{L} \otimes \ketbra{R} \e^{\frac{it\hat{H}}{\hbar}}, \\
\Phi_t(\ketbra{L} \otimes \ket{L}\bra{R}) &= \e^{- \frac{it\hat{H}}{\hbar}} \ketbra{L} \otimes \ket{L}\bra{R} \e^{\frac{it\hat{H}}{\hbar}}, \\
\Phi_t(\ketbra{L} \otimes \ket{R}\bra{L}) &= \e^{- \frac{it\hat{H}}{\hbar}} \ketbra{L} \otimes \ket{R}\bra{L} \e^{\frac{it\hat{H}}{\hbar}}, \\
\Phi_t(\ketbra{R} \otimes \ket{L}\bra{R}) &= \e^{- \frac{it\hat{H}}{\hbar}} \ketbra{R} \otimes \ket{L}\bra{R} \e^{\frac{it\hat{H}}{\hbar}}, \\
\Phi_t(\ketbra{R} \otimes \ket{R}\bra{L}) &= \e^{- \frac{it\hat{H}}{\hbar}} \ketbra{R} \otimes \ket{R}\bra{L} \e^{\frac{it\hat{H}}{\hbar}}. \label{eq:analytic-Phit-RRRL}
\end{align}

Now we will proceed by imposing the constraint that $\Phi_t$ has to be completely positive, we will do so by using the Choi matrix of $\Phi_t$ \cite{jamiolkowski1972linear,choi1975completely}. The Choi matrix of $\Phi_t$ is defined as
\begin{equation}
J(\Phi_t) = \sum_{k,\ell \in \{LL,LR,RL,RR\}} \Phi_t(\ket{k}\bra{\ell}) \otimes \ket{k}\bra{\ell}
\end{equation}
which we will write down as the block matrix
\begin{equation}
J(\Phi_t) = 
\begin{pmatrix}
\Phi_t(\ket{LL}\bra{LL}) & \Phi_t(\ket{LL}\bra{LR}) & \Phi_t(\ket{LL}\bra{RL}) & \Phi_t(\ket{LL}\bra{RR}) \\
\Phi_t(\ket{LR}\bra{LL}) & \Phi_t(\ket{LR}\bra{LR}) & \Phi_t(\ket{LR}\bra{RL}) & \Phi_t(\ket{LR}\bra{RR}) \\
\Phi_t(\ket{RL}\bra{LL}) & \Phi_t(\ket{RL}\bra{LR}) & \Phi_t(\ket{RL}\bra{RL}) & \Phi_t(\ket{RL}\bra{RR}) \\
\Phi_t(\ket{RR}\bra{LL}) & \Phi_t(\ket{RR}\bra{LR}) & \Phi_t(\ket{RR}\bra{RL}) & \Phi_t(\ket{RR}\bra{RR})
\end{pmatrix}.
\end{equation}
Then $\Phi_t$ is completely positive if and only if the Choi matrix $J(\Phi_t)$ is positive semidefinite, $J(\Phi_t) \geq 0$. Expressing all of the known terms as given by \eqref{eq:analytic-Phit-LL} - \eqref{eq:analytic-Phit-RR} and \eqref{eq:analytic-Phit-LRLL} - \eqref{eq:analytic-Phit-RRRL} yields
\begin{equation}
J(\Phi_t) = 
\begin{pmatrix}
\ket{LL}\bra{LL} & \e^{i (\varphi_{LL} - \varphi_{LR})} \ket{LL}\bra{LR} & \e^{i (\varphi_{LL} - \varphi_{RL})} \ket{LL}\bra{RL} & \Phi_t(\ket{LL}\bra{RR}) \\
\e^{i (\varphi_{LR} - \varphi_{LL})} \ket{LR}\bra{LL} & \ket{LR}\bra{LR} & \Phi_t(\ket{LR}\bra{RL}) & \e^{i (\varphi_{LR} - \varphi_{RR})} \ket{LR}\bra{RR} \\
\e^{i (\varphi_{RL} - \varphi_{LL})} \ket{RL}\bra{LL} & \Phi_t(\ket{RL}\bra{LR}) & \ket{RL}\bra{RL} & \e^{i (\varphi_{RL} - \varphi_{RR})} \ket{RL}\bra{RR} \\
\Phi_t(\ket{RR}\bra{LL}) & \e^{i (\varphi_{RR} - \varphi_{LR})} \ket{RR}\bra{LR} & \e^{i (\varphi_{RR} - \varphi_{RL})} \ket{RR}\bra{RL}) & \ket{RR}\bra{RR}
\end{pmatrix}.
\end{equation}
If $J(\Phi_t) \geq 0$, then also $\Pi J(\Phi_t) \Pi \geq 0$ for arbitrary self-adjoint operator $\Pi$. We will need only the case when $\Pi$ is a projection, in which case $\Pi J(\Phi_t) \Pi$ is a minor of $J(\Phi_t)$. For $\Pi = \I \otimes (\ketbra{LL} + \ketbra{RR})$ we get
\begin{equation}
\begin{pmatrix}
\ket{LL}\bra{LL} & \Phi_t(\ket{LL}\bra{RR}) \\
\Phi_t(\ket{RR}\bra{LL}) & \ket{RR}\bra{RR}
\end{pmatrix} \geq 0
\end{equation}
It now follows, for example by using \cite[Proposition 1.3.2]{bhatia2009positive} that we must have
\begin{align}
\Phi_t(\ket{LL}\bra{RR}) &= \alpha \ket{LL}\bra{RR} \\
\Phi_t(\ket{RR}\bra{LL}) &= \bar{\alpha} \ket{RR}\bra{LL},
\end{align}
where $\alpha \in \CC$ is a complex number and $\bar{\alpha}$ denotes the complex conjugate of $\alpha$. Similarly, for $\Pi = \I \otimes (\ketbra{LR} + \ketbra{RL})$ we get
\begin{equation}
J(\Phi_t) = 
\begin{pmatrix}
\ket{LR}\bra{LR} & \Phi_t(\ket{LR}\bra{RL}) \\
\Phi_t(\ket{RL}\bra{LR}) & \ket{RL}\bra{RL}
\end{pmatrix} \geq 0
\end{equation}
which now yields
\begin{align}
\Phi_t(\ket{LR}\bra{RL}) &= \beta \ket{LR}\bra{RL} \\
\Phi_t(\ket{RL}\bra{LR}) &= \bar{\beta} \ket{RL}\bra{LR}
\end{align}
for some $\beta \in \CC$. We thus get
\begin{equation}
J(\Phi_t) = 
\begin{pmatrix}
\ket{LL}\bra{LL} & \e^{i (\varphi_{LL} - \varphi_{LR})} \ket{LL}\bra{LR} & \e^{i (\varphi_{LL} - \varphi_{RL})} \ket{LL}\bra{RL} & \alpha \ket{LL}\bra{RR} \\
\e^{i (\varphi_{LR} - \varphi_{LL})} \ket{LR}\bra{LL} & \ket{LR}\bra{LR} & \beta \ket{LR}\bra{RL} & \e^{i (\varphi_{LR} - \varphi_{RR})} \ket{LR}\bra{RR} \\
\e^{i (\varphi_{RL} - \varphi_{LL})} \ket{RL}\bra{LL} & \bar{\beta}\ket{RL}\bra{LR} & \ket{RL}\bra{RL} & \e^{i (\varphi_{RL} - \varphi_{RR})} \ket{RL}\bra{RR} \\
\bar{\alpha} \ket{RR}\bra{LL} & \e^{i (\varphi_{RR} - \varphi_{LR})} \ket{RR}\bra{LR} & \e^{i (\varphi_{RR} - \varphi_{RL})} \ket{RR}\bra{RL} & \ket{RR}\bra{RR}
\end{pmatrix}.
\end{equation}
We now get that $J(\Phi_t) \geq 0$ if and only if
\begin{equation}
\tilde{J} =
\begin{pmatrix}
1 & \e^{i (\varphi_{LL} - \varphi_{LR})} & \e^{i (\varphi_{LL} - \varphi_{RL})} & \alpha \\
\e^{i (\varphi_{LR} - \varphi_{LL})} & 1 & \beta & \e^{i (\varphi_{LR} - \varphi_{RR})} \\
\e^{i (\varphi_{RL} - \varphi_{LL})} & \bar{\beta} & 1 & \e^{i (\varphi_{RL} - \varphi_{RR})} \\
\bar{\alpha} & \e^{i (\varphi_{RR} - \varphi_{LR})} & \e^{i (\varphi_{RR} - \varphi_{RL})} & 1
\end{pmatrix} \geq 0,
\end{equation}
we obtain this either by removing the columns and rows of $J(\Phi_t)$ that are identically zero, or by choosing $\Pi = \ketbra{LL} \otimes \ketbra{LL} + \ketbra{LR} \otimes \ketbra{LR} + \ketbra{RL} \otimes \ketbra{RL} + \ketbra{RR} \otimes \ketbra{RR}$. We will now use the condition that if $\tilde{J} \geq 0$, then the determinants of its minors must be non-negative, this follows by using that the minors must be positive semidefinite themselves and that the determinant is a product of the eigenvalues hence nonnegative. Considering the first minor of $\tilde{J}$ obtained by removing the last row and column, we get
\begin{equation}
\det
\begin{pmatrix}
1 & \e^{i (\varphi_{LL} - \varphi_{LR})} & \e^{i (\varphi_{LL} - \varphi_{RL})} \\
\e^{i (\varphi_{LR} - \varphi_{LL})} & 1 & \beta \\
\e^{i (\varphi_{RL} - \varphi_{LL})} & \bar{\beta} & 1
\end{pmatrix}
\geq 0
\end{equation}
which, using the Leibniz formula, yields
\begin{equation}
\beta \e^{i(\varphi_{RL} - \varphi_{LR})} + \bar{\beta} \e^{i(\varphi_{LR} - \varphi_{RL})} - \abs{\beta}^2 - 1 \geq 0
\end{equation}
which can be rewritten as
\begin{equation}
0 \geq (\beta - \e^{i(\varphi_{LR} - \varphi_{RL})})(\bar{\beta} - \e^{i(\varphi_{RL} - \varphi_{LR})}).
\end{equation}
Now since $\bar{\beta} - \e^{i(\varphi_{RL} - \varphi_{LR})}$ is the complex conjugate of $\beta - \e^{i(\varphi_{LR} - \varphi_{RL})}$, we must have $(\beta - \e^{i(\varphi_{LR} - \varphi_{RL})})(\bar{\beta} - \e^{i(\varphi_{RL} - \varphi_{LR})}) \geq 0$, which implies $(\beta - \e^{i(\varphi_{LR} - \varphi_{RL})})(\bar{\beta} - \e^{i(\varphi_{RL} - \varphi_{LR})}) = 0$ and $\beta = \e^{i(\varphi_{LR} - \varphi_{RL})}$ follows. We thus have
\begin{equation} \label{eq:analytic-Phit-LRRL}
\Phi_t(\ket{LR}\bra{RL}) = \e^{i(\varphi_{LR} - \varphi_{RL})} \ket{LR}\bra{RL} = \e^{- \frac{it\hat{H}}{\hbar}} \ket{LR}\bra{RL} \e^{\frac{it\hat{H}}{\hbar}}.
\end{equation}

Similarly, considering the first minor of $\tilde{J}$ obtained by dropping the second row and column we get
\begin{equation}
\det
\begin{pmatrix}
1 & \e^{i (\varphi_{LL} - \varphi_{RL})} & \alpha \\
\e^{i (\varphi_{RL} - \varphi_{LL})} & 1 & \e^{i (\varphi_{RL} - \varphi_{RR})} \\
\bar{\alpha} & \e^{i (\varphi_{RR} - \varphi_{RL})} & 1
\end{pmatrix}
\geq 0.
\end{equation}
Leibniz formula again yields
\begin{equation}
0 \geq \abs{\alpha}^2 - \alpha \e^{i(\varphi_{RR} - \varphi_{LL})} - \bar{\alpha} \e^{i(\varphi_{LL} - \varphi_{RR})} + 1 = (\alpha - \e^{i(\varphi_{LL} - \varphi_{RR})})(\bar{\alpha} - \e^{i(\varphi_{RR} - \varphi_{LL})})
\end{equation}
and we again get $\alpha = \e^{i(\varphi_{LL} - \varphi_{RR})}$ and
\begin{equation}
\Phi_t(\ket{LL}\bra{RR}) = \e^{i(\varphi_{LL} - \varphi_{RR})} \ket{LL}\bra{RR} = \e^{- \frac{it\hat{H}}{\hbar}} \ket{LL}\bra{RR} \e^{\frac{it\hat{H}}{\hbar}}.
\end{equation}
It thus follows that the Choi matrix of $\Phi_t$ and of the unitary channel given by $\e^{- \frac{it\hat{H}}{\hbar}}$ coincide, and thus for any $A \in \bound(\Ha_2 \otimes \Ha_2)$ we get
\begin{equation}
\Phi_t(A) = \e^{- \frac{it\hat{H}}{\hbar}} A \e^{\frac{it\hat{H}}{\hbar}}.
\end{equation}
In other words, we have just proved that, given \eqref{eq:analytic-Phit-psiYY} and \eqref{eq:analytic-Phit-XXpsi}, the time evolution of two delocalized particles is given by the Schr\"{o}dinger equation.

\section{Numerical proof that Schr\"{o}dinger equation for single delocalized system and positivity of time evolution implies gravity-mediated entanglement} \label{appendix:numeric}
A minimal assumption on $\Phi_t$ is that it is positive, that is for arbitrary $\rho \in \dens(\Ha_2 \otimes \Ha_2)$ we have
\begin{equation}
\Phi_t(\rho) \geq 0.
\end{equation}
This condition is hard to enforce for all $\rho \in \dens(\Ha_2 \otimes \Ha_2)$, but we can use semidefinite programming (SDP) \cite{boyd2004convex} to enforce this constraint for some subset of states. Thus let $\{ \rho_i \}_{i \in I} \subset \dens(\Ha_2 \otimes \Ha_2)$ be a selected set of states, then, in terms of the Choi matrix, we will require that
\begin{equation} \label{eq:numeric-pos}
\Tr_B(J(\Phi_t) (\I \otimes \rho_i)) \geq 0
\end{equation}
for all $i \in I$. The conditions \eqref{eq:analytic-Phit-LL} - \eqref{eq:analytic-Phit-RR} and
\begin{align}
\Phi_t(\ket{L}\bra{R} \otimes \ketbra{L}) &= \lambda_1 \e^{- \frac{it\hat{H}}{\hbar}} \ket{L}\bra{R} \otimes \ketbra{L} \e^{\frac{it\hat{H}}{\hbar}}, \\
\Phi_t(\ket{R}\bra{L} \otimes \ketbra{L}) &= \lambda_1 \e^{- \frac{it\hat{H}}{\hbar}} \ket{R}\bra{L} \otimes \ketbra{L} \e^{\frac{it\hat{H}}{\hbar}}, \\
\Phi_t(\ket{L}\bra{R} \otimes \ketbra{R}) &= \lambda_2 \e^{- \frac{it\hat{H}}{\hbar}} \ket{L}\bra{R} \otimes \ketbra{R} \e^{\frac{it\hat{H}}{\hbar}}, \\
\Phi_t(\ket{R}\bra{L} \otimes \ketbra{R}) &= \lambda_2 \e^{- \frac{it\hat{H}}{\hbar}} \ket{R}\bra{L} \otimes \ketbra{R} \e^{\frac{it\hat{H}}{\hbar}}, \\
\Phi_t(\ketbra{L} \otimes \ket{L}\bra{R}) &= \lambda_3 \e^{- \frac{it\hat{H}}{\hbar}} \ketbra{L} \otimes \ket{L}\bra{R} \e^{\frac{it\hat{H}}{\hbar}}, \\
\Phi_t(\ketbra{L} \otimes \ket{R}\bra{L}) &= \lambda_3 \e^{- \frac{it\hat{H}}{\hbar}} \ketbra{L} \otimes \ket{R}\bra{L} \e^{\frac{it\hat{H}}{\hbar}}, \\
\Phi_t(\ketbra{R} \otimes \ket{L}\bra{R}) &= \lambda_4 \e^{- \frac{it\hat{H}}{\hbar}} \ketbra{R} \otimes \ket{L}\bra{R} \e^{\frac{it\hat{H}}{\hbar}}, \\
\Phi_t(\ketbra{R} \otimes \ket{R}\bra{L}) &= \lambda_4 \e^{- \frac{it\hat{H}}{\hbar}} \ketbra{R} \otimes \ket{R}\bra{L} \e^{\frac{it\hat{H}}{\hbar}},
\end{align}
where $\mu_j \geq \lambda_0$ for all $j \in \{1, \ldots, 4\}$ and $\lambda_0$ is the upped bound on decoherence that comes from the experimental verification of the Schr\"{o}dinger equation, will be enforced using the identity
\begin{equation}
\Phi_t(\rho) = \Tr_B(J(\Phi_t) (\I \otimes \rho^T))
\end{equation}
where $\rho^T$ is the transpose of $\rho$. Finally trace preserving of $\Phi_t$ is captured by the condition
\begin{equation}
\Tr_A(J(\Phi_t)) = \I.
\end{equation}
Here we use $\Tr_A$ and $\Tr_B$ to denote the partial trace over the first two and last two Hilbert spaces respectively. These conditions enable us to optimize over an outer approximation of the set of positive time evolutions which coincide with the solution of the Schr\"{o}dinger equation for single delocalized system.

Finally we want to see whether the final state of the system can be separable. The initial state of the system in the proposed experiments consists of both interferometers being in the equal superposition of left and right arms, that is,
\begin{equation}
\ket{\psi(0)} = \ket{+} \otimes \ket{+} = \frac{1}{2} (\ket{LL} + \ket{LR} + \ket{RL} + \ket{RR}),
\end{equation}
and the final state is
\begin{equation}
\Phi_t(\ketbra{\psi(0)}) = \Tr_B(J(\Phi_t) (\I \otimes \ketbra{\psi(0)})).
\end{equation}
We have omitted the transpose since $\ketbra{\psi(0)}^T = \ketbra{\psi(0)}$. Since we are dealing with a state of two qubits, $\Phi_t(\ketbra{\psi(0)})$ is separable if and only if it has positive partial transpose \cite{peres1996separability,horodecki2001separability,fawzi2021set}, i.e., $\Phi_t(\ketbra{\psi(0)})^{T_1} \geq 0$. Here $T_1$ denotes the partial transpose over the first Hilbert space in the output of $\Phi_t$. Hence we will maximize over all numbers $\mu$ such that $\Phi_t(\ketbra{\psi(0)})^{T_1} \geq \mu \I$; if the maximum is negative, then the final state $\Phi_t(\ketbra{\psi(0)})$ is entangled for all positive time evolutions and for the given value of $\lambda_0$. We thus want to solve the following optimization problem in which we replace $J(\Phi_t)$ by an unknown variable $X \in \bound(\Ha_2^{\otimes 4})$:
\begin{equation}
\begin{split}
\max_{X \in \bound(\Ha_2^{\otimes 4})} \quad & \mu \\
\text{such that} \quad & \Tr_A(X) = \I \\
& \Tr_B(X (\I \otimes \rho_i)) \geq 0, \ \forall i \in I \\
& \Tr_B(X (\I \otimes \ketbra{LL})) = \ketbra{LL} \\
& \Tr_B(X (\I \otimes \ketbra{LR})) = \ketbra{LR} \\
& \Tr_B(X (\I \otimes \ketbra{RL})) = \ketbra{RL} \\
& \Tr_B(X (\I \otimes \ketbra{RR})) = \ketbra{RR} \\
& \Tr_B(X (\I \otimes \ket{R}\bra{L} \otimes \ketbra{L})) = \lambda_1 \e^{- \frac{it\hat{H}}{\hbar}} \ket{L}\bra{R} \otimes \ketbra{L} \e^{\frac{it\hat{H}}{\hbar}} \\
& \Tr_B(X (\I \otimes \ket{L}\bra{R} \otimes \ketbra{L})) = \lambda_1 \e^{- \frac{it\hat{H}}{\hbar}} \ket{R}\bra{L} \otimes \ketbra{L} \e^{\frac{it\hat{H}}{\hbar}} \\
& \Tr_B(X (\I \otimes \ket{R}\bra{L} \otimes \ketbra{R})) = \lambda_2 \e^{- \frac{it\hat{H}}{\hbar}} \ket{L}\bra{R} \otimes \ketbra{R} \e^{\frac{it\hat{H}}{\hbar}} \\
& \Tr_B(X (\I \otimes \ket{L}\bra{R} \otimes \ketbra{R})) = \lambda_2 \e^{- \frac{it\hat{H}}{\hbar}} \ket{R}\bra{L} \otimes \ketbra{R} \e^{\frac{it\hat{H}}{\hbar}} \\
& \Tr_B(X (\I \otimes \ketbra{L} \otimes \ket{R}\bra{L})) = \lambda_3 \e^{- \frac{it\hat{H}}{\hbar}} \ketbra{L} \otimes \ket{L}\bra{R} \e^{\frac{it\hat{H}}{\hbar}} \\
& \Tr_B(X (\I \otimes \ketbra{L} \otimes \ket{L}\bra{R})) = \lambda_3 \e^{- \frac{it\hat{H}}{\hbar}} \ketbra{L} \otimes \ket{R}\bra{L} \e^{\frac{it\hat{H}}{\hbar}} \\
& \Tr_B(X (\I \otimes \ketbra{R} \otimes \ket{R}\bra{L})) = \lambda_4 \e^{- \frac{it\hat{H}}{\hbar}} \ketbra{R} \otimes \ket{L}\bra{R} \e^{\frac{it\hat{H}}{\hbar}} \\
& \Tr_B(X (\I \otimes \ketbra{R} \otimes \ket{L}\bra{R})) = \lambda_4 \e^{- \frac{it\hat{H}}{\hbar}} \ketbra{R} \otimes \ket{R}\bra{L} \e^{\frac{it\hat{H}}{\hbar}} \\
& \lambda_j \geq \lambda_0, \ \forall j \in \{1, \ldots, 4\} \\
& (\Tr_B(X (\I \otimes \ketbra{\psi(0)})))^{T_1} \geq \mu \I
\end{split}
\end{equation}
The implementation of this semidefinite program uses the following parameters: both interferometers have the path separation of $\Delta x = 250 \unit{\mu m}$ in line with the original GME proposal \cite{bose2017spin}, interaction time is $t = 1 \unit{s}$, and the mass of the mesoscopic particles is $m_\gme = 1.485 \cdot 10^{-12} \unit{kg}$ which, as explained in the main text, is due to the assumed correspondence between the experiment verifying the Schr\"{o}dinger equation and the GME experiment. The center of mass distance is $d \approx 5.3115 \unit{cm}$, which is the sum of the radius of the $10 \unit{kg}$ Tungsten sphere $R \approx 4.9865 \unit{cm}$, the surface separation of $3 \unit{mm}$ coming from the respective experiment to verify the Schr\"{o}dinger equation, and the path separation $\Delta x = 250 \unit{\mu m}$. The set $\{ \rho_i \}_{i \in I}$ consisted of $1000$ randomly sampled pure states.

\end{document}